\documentclass[twocolumn,floatfix,prb,superscriptaddress,longbibliography]{revtex4-1}

\usepackage[english]{babel}
\usepackage{graphicx, float}
\usepackage{dcolumn}
\usepackage{bm}
\usepackage{lipsum}
\usepackage{mathtools}
\usepackage{braket,slashed}
\usepackage[caption=false]{subfig}
\usepackage{xcolor}
\definecolor{RED}{HTML}{fc0317}
\usepackage{amssymb}
\usepackage{soul}

\usepackage{hyperref}

\renewcommand{\selectlanguage}[1]{}
\begin{document}


\title{Emergence of multiple zero modes bound to vortices\\ in extended topological Josephson junctions}

\author{Adrian Reich} \email{adrian.reich@kit.edu}
\affiliation{Institute for Theoretical Condensed Matter Physics, Karlsruhe Institute of Technology, 76131 Karlsruhe, Germany}

\author{Kiryl Piasotski}
\affiliation{Institute for Theoretical Condensed Matter Physics, Karlsruhe Institute of Technology, 76131 Karlsruhe, Germany}
\affiliation{Institute for Quantum Materials and Technologies, Karlsruhe Institute of Technology, 76344 Eggenstein-Leopoldshafen, Germany}

\author{Eytan Grosfeld}
\affiliation{Department of Physics, Ben-Gurion University of the Negev, Beer-Sheva 8410501, Israel}

\author{Alexander Shnirman}
\affiliation{Institute for Theoretical Condensed Matter Physics, Karlsruhe Institute of Technology, 76131 Karlsruhe, Germany} 
\affiliation{Institute for Quantum Materials and Technologies, Karlsruhe Institute of Technology, 76344 Eggenstein-Leopoldshafen, Germany}

\date{\today}

\begin{abstract}

We study planar Josephson junctions formed on the surface of a three-dimensional topological insulator (Fu-Kane proposal). We
examine the experimentally relevant parameter regimes in which the frequently used effective description in terms of two counter-propagating one-dimensional Majorana modes with hybridization dependent on the Josephson phase difference reaches its validity limit. This happens when the effective velocity of the emergent one-dimensional Majorana modes approaches zero. As parameters like the chemical potential or the width of the junction are tuned, instances of vanishing effective velocity mark the emergence of additional `Dirac cones' at zero energy and finite momentum. If the junction is subjected to an external magnetic field, Josephson vortices may then bind a number of zero modes in addition to the topological Majorana mode. The additional zero modes are `symmetry-protected' and can be lifted by a broken mirror symmetry (which is to be expected in realistic scenarios) as well as by an in-plane magnetization (or Zeeman field). We note that the ensuing presence of additional low-energy Andreev states can significantly contribute to measured quantities like the Josephson current or microwave absorption spectra.

\end{abstract}

\maketitle


\section{Introduction}

Hybrid structures built from topological insulators and superconductors\cite{fu_superconducting_2008,fu_probing_2009} (Fu-Kane proposal) have emerged as a prominent platform in the ongoing search for Majorana zero modes\cite{volovik1999fermion,read_paired_2000,ivanov2001nonabelian,Kitaev_2001}. When superconducting order is induced in the spin-momentum-locked surface states of a three-dimensional strong topological insulator, the resulting system is expected to host effective topological superconductivity and localized Majorana states at boundaries and in vortices. This prospect has driven sustained theoretical\cite{law_majorana2009,tanaka_manipulation_2009,cook_majorana2011,ioselevich_anomalous_2011,potter_anomalous_2013,park2015detecting,choi_josephson_2019,hegde_topological_2020,park2020electron,backens_current--phase_2021,backens_topological_2022,piasotski_topological_2024,laubscher2024detectionmajoranazeromodes} and experimental efforts\cite{williams2012unconventional,veldhorst2012josephson,kurter2015evidence,sochnikov2015nonsinusoidal,wiedenmann_4-periodic_2016,kayyalha2020highly,yue_signatures_2024,zhang2022ac,park2024corbino,park2026vortexparitycontrolleddiodeeffectcorbino}, establishing these hetero-structures as a focal point in the broader exploration of non-Abelian anyonic quasiparticles and their potential role in fault-tolerant quantum information processing\cite{chetan2008nonabelian}.

Within the general proposal of emergent two-dimensional  topological superconductivity,
extended planar Josephson junctions on the surface of topological insulators have first been proposed and analyzed in a seminal paper by Fu and Kane \cite{fu_superconducting_2008}. They have derived an effective low-energy theory in terms of one-dimensional Majorana modes counter-propagating along the junction. This `$k\cdot p$'-approximation, which is what we will in the following sometimes refer to as `Fu-Kane theory', has proven extremely insightful and formed the basis for numerous theoretical predictions and interpretations of experimental data\cite{potter_anomalous_2013,choi_josephson_2019,hegde_topological_2020,backens_current--phase_2021,backens_topological_2022,yue_signatures_2024,piasotski_topological_2024}. If such a system is placed in an external magnetic field, one of the basic expectations is the presence of a spectrum of bound states (a.k.a. Caroli-de Gennes-Matricon (CdGM) states~\cite{CAROLI1964307}) localized in the Josephson vortices with energies $\pm E_n \propto \sqrt{n}, n=0,1,2,\dots$, including a single Majorana bound state at zero energy\cite{grosfeld_observing_2011,potter_anomalous_2013}.

In this work, we adopt the Fu-Kane 2D continuum model of the surface states, which is valid for single particle energies $\xi$ (counted from the appropriate chemical potential) well within the bulk band gap $E_g$ of the 3D topological insulator, $|\xi| < E_g$, and demonstrate that the 1D Fu-Kane effective description of the topological Josephson junction, derived for energies within the induced superconducting gap $\Delta_0$, i.e., $|\xi|<\Delta_0 \ll E_g$, loses its validity as the effective velocity of the one-dimensional Majorana edge modes tends to zero. The existence of such a regime is, of course, not surprising, as any effective theory based on a low energy/momentum expansion has a limited domain of validity. Concretely, the `$k\cdot p$'-method misses additional low energy localized states, which manifest themselves as zero modes in an ideal situation and near-zero modes in experimentally realistic scenarios. We argue that these additional CdGM states might play an important role in the interpretation of recently conducted and proposed experiments. 

\section{Topological Josephson junction in a transverse magnetic field}

\begin{figure}
    \centering
    \includegraphics[width=0.8\linewidth]{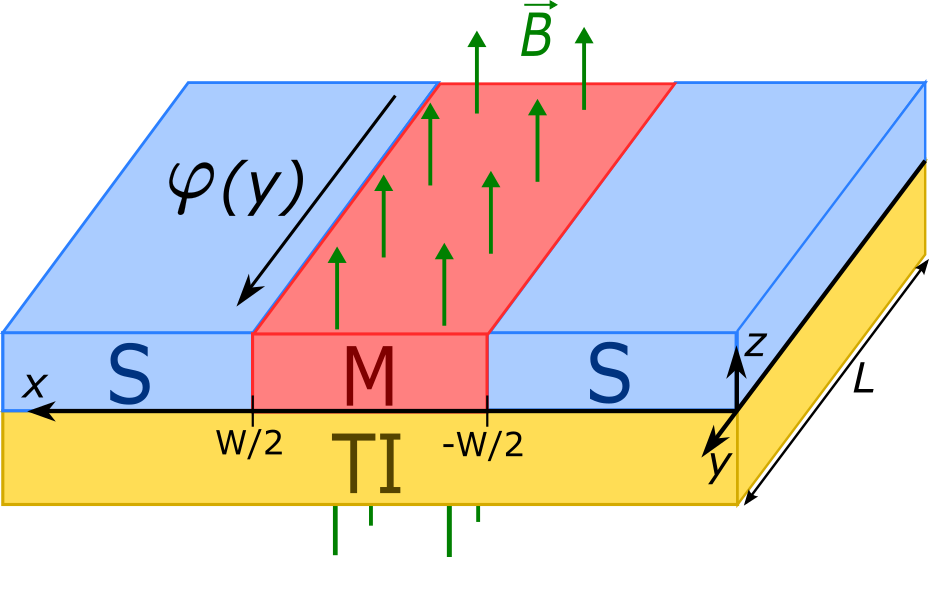}
    \includegraphics[width=.9\linewidth]{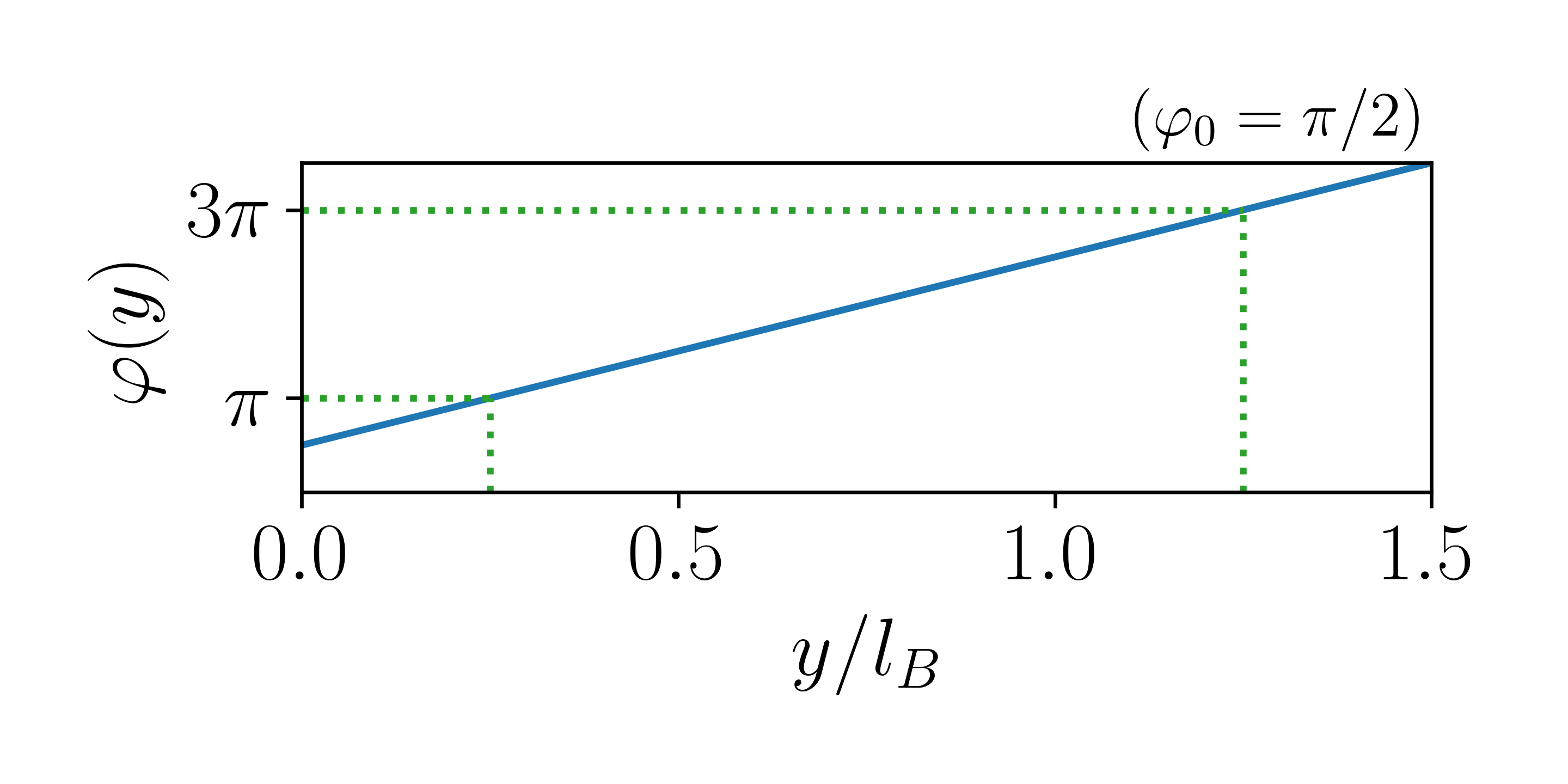}
    \caption{Sketch of the superconductor-ferromagnet-superconductor (SMS) junction on the surface a three-dimensional topological insulator (TI) with width $W$ and length $L$. The external magnetic field $\vec{B} = B\hat{e}_z$ leads to a running phase difference $\varphi(y) = 2\pi y/l_B+\varphi_0$ (sketched below) between the superconductors with magnetic length $l_B \propto 1/B$. The dotted green lines indicate the positions of Josephson vortices, separated by a distance of $l_B$.}
    \label{fig:sketchSMSinmagneticfield}
\end{figure}

We consider a Josephson junction formed between two $s$-wave superconductors on the surface of three-dimensional strong topological insulator (TI) in an external magnetic field $\Vec{B} = \vec\nabla\times\Vec{A}$ perpendicular to the surface (Fig.~\ref{fig:sketchSMSinmagneticfield}). We assume the superconducting gaps to be of the same magnitude $\Delta_0$ on both sides of the junction, but differing by a relative phase $\varphi(y)$. The position dependence of the phase difference is a result of the magnetic flux.~\cite{Barone-Josephson} Additionally, we take into account the effect of a ferromagnetic insulator deposited between the superconductors. Its magnetization causes an effective exchange field $\vec{M} = (M_x,M_y,M_z)^T$ in the underlying surface of the TI which couples to the electrons' spin. Though ferromagnet-superconductor interfaces on the surface of TIs have to the best of our knowledge not yet been experimentally realized, we include these terms out of academic interest, in accordance with some of the authors' earlier theoretical works\cite{backens_current--phase_2021,reich_magnetization_2023,reich_berezinskii-kosterlitz-thouless_2024}, in order to understand their effect on the phenomena we study below. For comparison with current experiments one should assume $\vec M=0$.

The corresponding Bogoliubov-de Gennes (BdG) Hamiltonian reads
\begin{align} \label{eq:hBdG-inmagneticfield}
    &h(x,y) = \\
    &\Big[-iv_F\partial_x\sigma_x+\left(-iv_F\partial_y+\frac{ev_F}{c}A_y(x)\tau_z\right)\sigma_y- \mu(x)\Big]\tau_z \notag\\ &\quad+ \vec{M}(x)\cdot\vec\sigma
    + \Delta_0(x)\left(\cos\varphi(x,y)\tau_x + \sin\varphi(x,y)\tau_y\right) \notag
\end{align}
where 
$\Delta_0(x) = \Delta_0\,\theta(|x|-W/2)$, $\varphi(x,y) = \varphi(y)\,\theta(x)$, $\vec{M}(x) = \vec{M}\,\theta(W/2-|x|)$, $\mu(x) = \mu_N\,\theta(W/2-|x|) + \mu_S\,\theta(|x|-W/2)$.
$\tau_i$ and $\sigma_i$ are Pauli matrices in Nambu space and spin space, respectively. $W$ is the width of the junction and $v_F$ the surface state Fermi velocity.

To account for a possible shift (pinning) of the chemical potential $\mu$ due to proximity to the metallic superconductors, we allow it to exhibit a step-like profile in $x$-direction. The experimentally relevant case, from which we will draw the examples below, corresponds to $\mu_S$ constituting the largest energy scale in the model (`Andreev limit'). The chemical potential $\mu_N$ in the non-superconducting region can in principle be controlled by a gate voltage and will therefore be chosen as a tuning parameter. 

The orbital effects of the applied magnetic field are accounted for by the vector potential $\Vec{A}=(0,A_y(x),0)^T$ in Landau gauge, which, under the assumption of London screening with London penetration depth $\lambda_L$, is given by
\begin{align}
\frac{e}{c}A_y&(x) = \frac{\pi}{l_B W}\times\\
&\times\begin{cases}
    \lambda_L e^{(x+\frac{W}{2})/\lambda_L},\quad &x<-W/2, \\
    \left(x+\frac{W}{2}+\lambda_L\right),\quad &|x|\leq W/2,\\
    \left(W+\lambda_L\left(2-e^{-(x-\frac{W}{2})/\lambda_L}\right)\right),\quad &x>W/2. \label{eq:Ay(x)}
\end{cases}\notag
\end{align}
The 1D magnetic length $l_B\equiv \frac{\Phi_0}{(2\lambda_L+W)B}$ is inversely proportional to the strength of the applied magnetic field. Under the assumption $l_B\ll \lambda_J$, where $\lambda_J$ is the Josephson penetration depth of the junction, the phase difference then grows linearly
\begin{align}
    \varphi(y) = \frac{2\pi}{l_B}y + \varphi_0
\end{align}
with some phase offset $\varphi_0$. Majorana zero modes as well as low-energy Andreev states are localized at points $y_m$ at which $\varphi(y_m) = (2m+1)\pi$, $m\in\mathbb{Z}$, in the center of Josephson vortices,~\cite{grosfeld_observing_2011,potter_anomalous_2013} which are thus separated by a distance $l_B$.

\section{The Fu-Kane effective theory\\ and its range of validity}
\subsection{Low-energy approximation and its applications}
The effective low-energy theory based on the groundbreaking insights by Fu and Kane~\cite{fu_superconducting_2008,fu_probing_2009}, in which the topological insulator Josephson junction is described by means of a 2$\times$2 Hamiltonian in one spatial dimension, has in the past been utilized in a variety of works, e.g. examining experimental signatures of Majorana modes in the form of the (anomalous) Josephson current~\cite{potter_anomalous_2013,backens_current--phase_2021}, lifting of zeros in the Fraunhofer diffraction pattern~\cite{hegde_topological_2020,yue_signatures_2024,laubscher2024detectionmajoranazeromodes} as well as braiding operations~\cite{choi_josephson_2019,hegde_topological_2020}.

The main idea can be sketched as follows (where we begin under the assumption $M_x=M_y=0$ and defer the discussion of the effects of an in-plane magnetization to Sec.~\ref{sec:translationallyinvariant}). Starting from a 4$\times$4 BdG Hamiltonian of the form \eqref{eq:hBdG-inmagneticfield}, describing two-dimensional TI surface states, a projection onto the 2$\times$2 subspace of linearly dispersing one-dimensional Majorana modes which are bound to the junction is carried out, yielding a Hamiltonian of the form
\begin{align}
    h_\text{eff} = -iv\partial_y\rho_x + \varepsilon(y)\rho_z
\end{align}
with an effective velocity $v$, provided that the effective mass term $\varepsilon(y)$ only varies slowly with the coordinate $y$ along the junction. $\rho_i$ are here Pauli matrices acting in the low-energy subspace. The effective velocity arises as a correction from the $-i\partial_y$ contribution to the original Hamiltonian, in which all parameters are initially treated as effectively $y$-independent.

In our example, the physical quantity responsible for the mass term $\varepsilon(y)$ is the phase difference $\varphi(y)$ between the superconductors.~\cite{fu_superconducting_2008} Specifically, it holds $\varepsilon(y)/\Delta_0 \sim \pi-\varphi(y)$. Points where $\varepsilon(y)$ changes sign then correspond to Josephson vortices~\cite{Barone-Josephson}, which ensue the existence of zero-dimensional bound states in their center. Particularly, a topologically protected zero-energy solution, i.e. a Majorana bound state, is guaranteed. Depending on the details, a number of additional localized Andreev bound states with finite energies may be present as well.~\cite{grosfeld_observing_2011,potter_anomalous_2013}

Recently, the treatment has been generalized to more carefully take into account the $y$-dependence of the model parameters~\cite{backens_topological_2022,piasotski_topological_2024} and thereby arrive at a Hamiltonian of the form
\begin{align}
    h_\text{eff} = -\frac{i}{2}\{v(y),\partial_y\}\rho_x + \varepsilon(y)\rho_z, \label{eq:heff_anticomm}
\end{align}
involving an effective velocity $v \rightarrow v(y)$ which depends on the position along the junction as well. The constant effective velocity from above corresponds to taking $v(y_0)$ at a point $y_0$ where $\varepsilon(y_0)=0$, i.e. in our model at $\varphi = \pi$. In Ref.~\onlinecite{piasotski_topological_2024}, it has been proposed that this spatial dependence of $v$ in combination with irregularities of the fabricated junctions can explain the anomalous Josephson currents experimentally observed in Refs.~\onlinecite{zhang2022ac,park2024corbino}.

Note that the effective velocity is an oscillating quantity as a function of some system parameters like the chemical potential $\mu$ and the width of the junction $W$. E.g. for $\mu \equiv \mu_N=\mu_S$ and $\mu\gg\Delta_0$, Fu and Kane~\cite{fu_superconducting_2008} found for the S-TI-S junction $v \propto \cos(\mu W/v_F)$. In a similar calculation for ballistic graphene in the regime $\mu_S\gg \Delta_0,\mu_N$, Titov and Beenakker~\cite{titov_josephson_2006} obtained $v \propto \sin(\mu_N W/v_F)$. In Ref.~\onlinecite{piasotski_topological_2024}, a general formula which contains both of these special cases has been derived. Realistic systems thus might feature points at which the effective velocity vanishes, $v\rightarrow 0$. 

\subsection{Singularities near zero-velocity points}
In the vicinity of such a zero-velocity point, take $v(y) \approx a y$ with some slope $a$, and $\varepsilon(y) \simeq \varepsilon_0$ to be approximately constant in the effective Hamiltonian \eqref{eq:heff_anticomm}. One finds
\begin{align}
    h_\text{eff}^2 = -a^2y^2\partial_y^2-2a^2y\partial_y+(\varepsilon_0^2-a^2/4).
\end{align}
The general solution to $h_\text{eff}^2\psi = E^2\psi$ (which can e.g. be found by means of the Frobenius method \cite{teschl2012ordinary}) reads
\begin{align}
    \psi(y) = c_1 y^{-\frac{1}{2}-\sqrt{\varepsilon_0^2-E^2}/a} + c_2 y^{-\frac{1}{2}+\sqrt{\varepsilon_0^2-E^2}/a}.
\end{align}
A solution which is normalizable thus only exists as long as $E^2<\varepsilon_0^2$, with the wave function diverging at $y=0$ if $E^2<\varepsilon_0^2-a^2/4$ is not met.

The Hamiltonian \eqref{eq:heff_anticomm} can in fact be mapped to a Dirac Hamiltonian in a 1+1-dimensional curved space-time with a spatially varying mass $m(y) \sim \varepsilon(y)/v(y)$ and metric $ds^2 = v^2(y)dt^2-dy^2$. The points at which the effective velocity vanishes $v(y) = 0$ then correspond to the event horizons of Schwarzschild black holes, effectively slicing the system into two `universes' between which no information can be passed~\cite{Mann_Semiclassical1991}. This singularity ensues the observed divergences of the wave functions.

These peculiar results indicate that the zero-velocity points mark the edge of the validity of the employed approximations since they can be attributed to strictly limiting the theory to the low-energy subspace. In Appendix \ref{app:higherordercorrections}, we show that taking into account higher order corrections to the low-energy approximation, corresponding to virtual processes involving the quasiparticle continuum, the singular behavior is regularized as a contribution $\sim \partial_y^2\rho_z$ to $h_\text{eff}$ is generated (among other additional terms).

It should be noted that alternative numerical~\cite{laubscher2024detectionmajoranazeromodes} and analytical~\cite{park2015detecting}  approaches within the Fu-Kane proposal, which, in principle, are capable of going beyond the low-energy `$k\cdot p$'-approximation, have been introduced. We notice also a closely related 2D tight-binding model studied in Ref.~\cite{abboud_signatures_2022}.

\subsection{Emergence of additional low-energy degrees of freedom in the translationally invariant case \label{sec:translationallyinvariant}}

In the case of no external magnetic field and thus translational symmetry in $y$-direction, the momentum $p_y \equiv k$ is conserved and the problem can be solved for each value of $k$ separately. Instead of taking $k=0$ and subsequently including small values of $k$ perturbatively, as has been the strategy in Ref.~\onlinecite{fu_superconducting_2008}, let us here thus find the low-energy eigenmodes of
\begin{align}
    h_k(x) = &\left[-iv_F\partial_x\sigma_x+v_Fk\sigma_y - \mu(x)\right]\tau_z \label{eq:fukanehamiltonianwithk} \\
    &+\vec{M}(x)\cdot\vec{\sigma} + \Delta_0(x)\left(\cos\varphi(x)\tau_x + \sin\varphi(x)\tau_y\right), \notag 
\end{align}
directly for each value of $k$, with $\mu(x)$, $\vec{M}(x)$, $\Delta_0(x)$ and $\varphi(x)$ as given below Eq.~\eqref{eq:hBdG-inmagneticfield}.

The Hamiltonian \eqref{eq:fukanehamiltonianwithk} possesses particle-hole symmetry $U_Ch_kU_C^{-1}=-h_{-k}^\ast$ with $U_C = \tau_y\sigma_y$. 

As long as $M_y=0$, there is furthermore a quasi-time-reversal symmetry $\Tilde{U}_Th_k\tilde{U}_T^{-1} = h_{-k}^\ast$ with $\tilde{U}_T = I_x e^{i\varphi\tau_z/2}$, where $I_x$ denotes $x$-inversion $I_x f(x)= f(-x)$. Note that $U_C$ and $\tilde{U}_T$ are also symmetries of the full Hamiltonian \eqref{eq:hBdG-inmagneticfield} in a magnetic field, though $\tilde{U}_T\rightarrow \tilde{U}_T(y)$ becomes a local operator due to the $y$-dependence of $\varphi$.

Since $(U_C\mathcal{K})^2 = (\tilde{U}_T\mathcal{K})^2 = 1$, where $\mathcal{K}$ denotes complex conjugation, the effective one-dimensional system belongs to the symmetry class BDI in the tenfold classification\cite{tenfoldclassification} and is characterized by a $\mathbb{Z}$-valued topological invariant. If $\tilde{U}_T$ does not hold due to a broken mirror symmetry (e.g. $\mu_S(x>0)\neq\mu_S(x<0)$) or $M_y\neq 0$, the system instead falls into the symmetry class $D$ and only a $\mathbb{Z}_2$ invariant remains. In Ref.~\onlinecite{pientka_topological_2017} the situation is very similar, though the dispersion there is quadratic with Rashba spin-orbit coupling.

\begin{figure*}
    \centering
    \includegraphics[width=\linewidth]{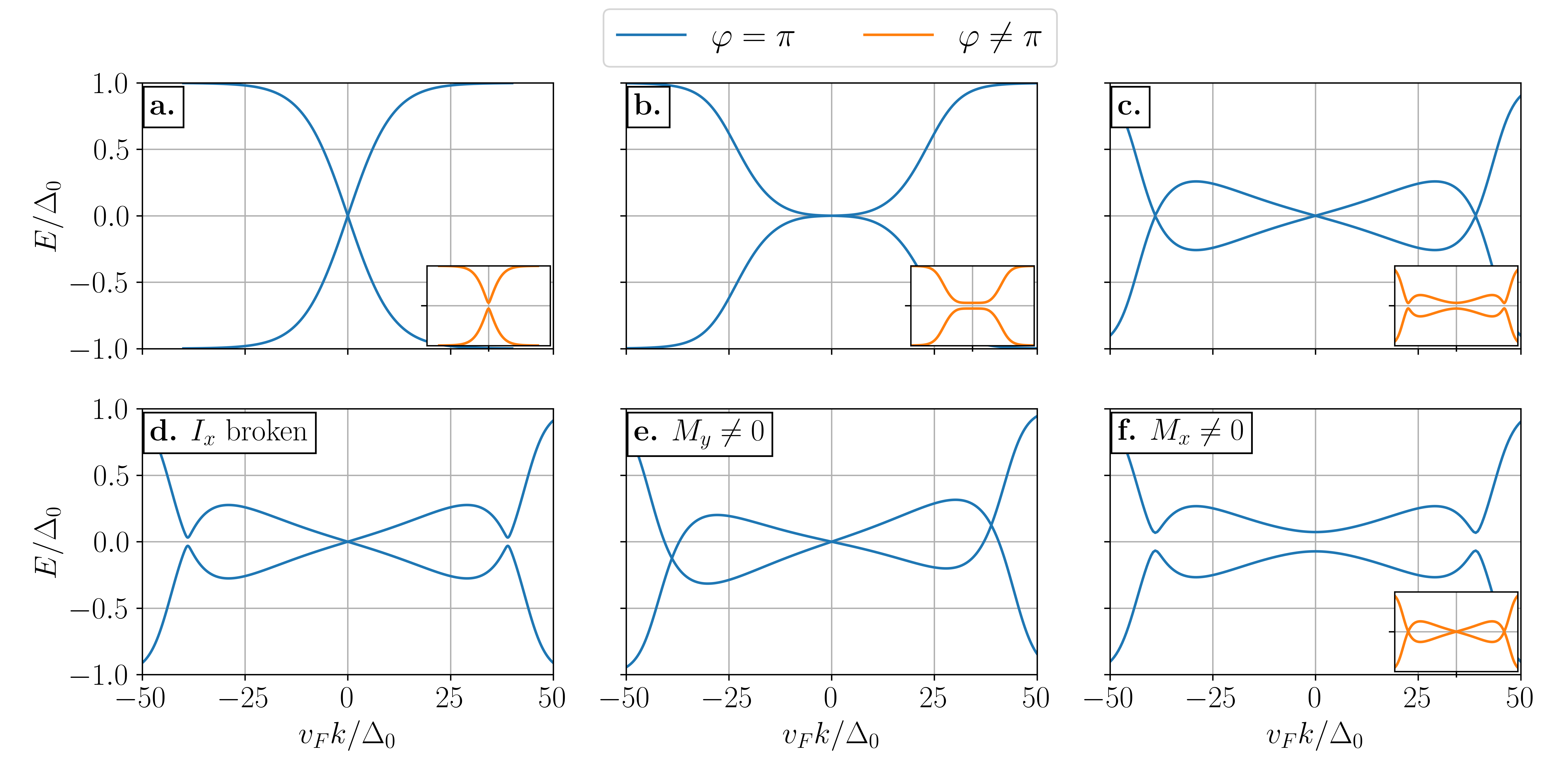}
    \caption{Dispersion of the one-dimensional bound states following from the Hamiltonian \eqref{eq:fukanehamiltonianwithk} for $\varphi=\pi$ (blue) and $\varphi=1.05\pi$ (orange in the insets) with $W=0.1\,v_F/\Delta_0$, $\mu_S = 100\Delta_0$ and (a) $\mu_N = 0$,  (b) $\mu_N = 31\Delta_0 \approx \mu_{N,1}^{(0)}$, (c)-(f) $\mu_N = 50\Delta_0$. Additionally in panel (d) an asymmetry has been introduced to the junction $\mu_S(x>W/2) = 1.2\mu_S(x<-W/2)$. In the panels (a)-(d) it holds $\vec{M}=0$, whereas for (e) ($M_y=\Delta_0$) and (f) ($M_x = 0.8\Delta_0$), an in-plane magnetization has been added. In (b) at $\mu_N\approx \mu_{N,1}^{(0)}$, where the effective velocity (i.e. the slope at $k=0$) becomes zero, the non-linear corrections to the previously linear dispersion (shown in (a), where the naive effective theory is still an adequate description) become apparent. Increasing $\mu_N$ further to what is shown in (c), two additional Dirac cones, i.e. linearly dispersing low-energy modes, at finite momenta appear. For $\varphi\neq \pi$ a gap opens in each case. The asymmetry in panel (d) results in gaps at the two additional Dirac points (but not at $k=0$) even for $\varphi=\pi$. A finite value of $M_y$ `tilts' the picture as shown in (e), such that the Dirac cones at finite $k$ are no longer centered at $E=0$. It is here interesting to note that, due to this tilting, sufficiently large values of $M_y$ may change the nature of the low-$k$ modes from counter- to co-propagating.\cite{donis2021chirality} A finite value of $M_x$ on the other hand, shown in (f), only acts like an additional phase difference and results in the gap closings being shifted to a value $\varphi\neq\pi$.}
    \label{fig:Dirac-cones}
\end{figure*}

\begin{figure}
    \centering
    \includegraphics[width=\linewidth]{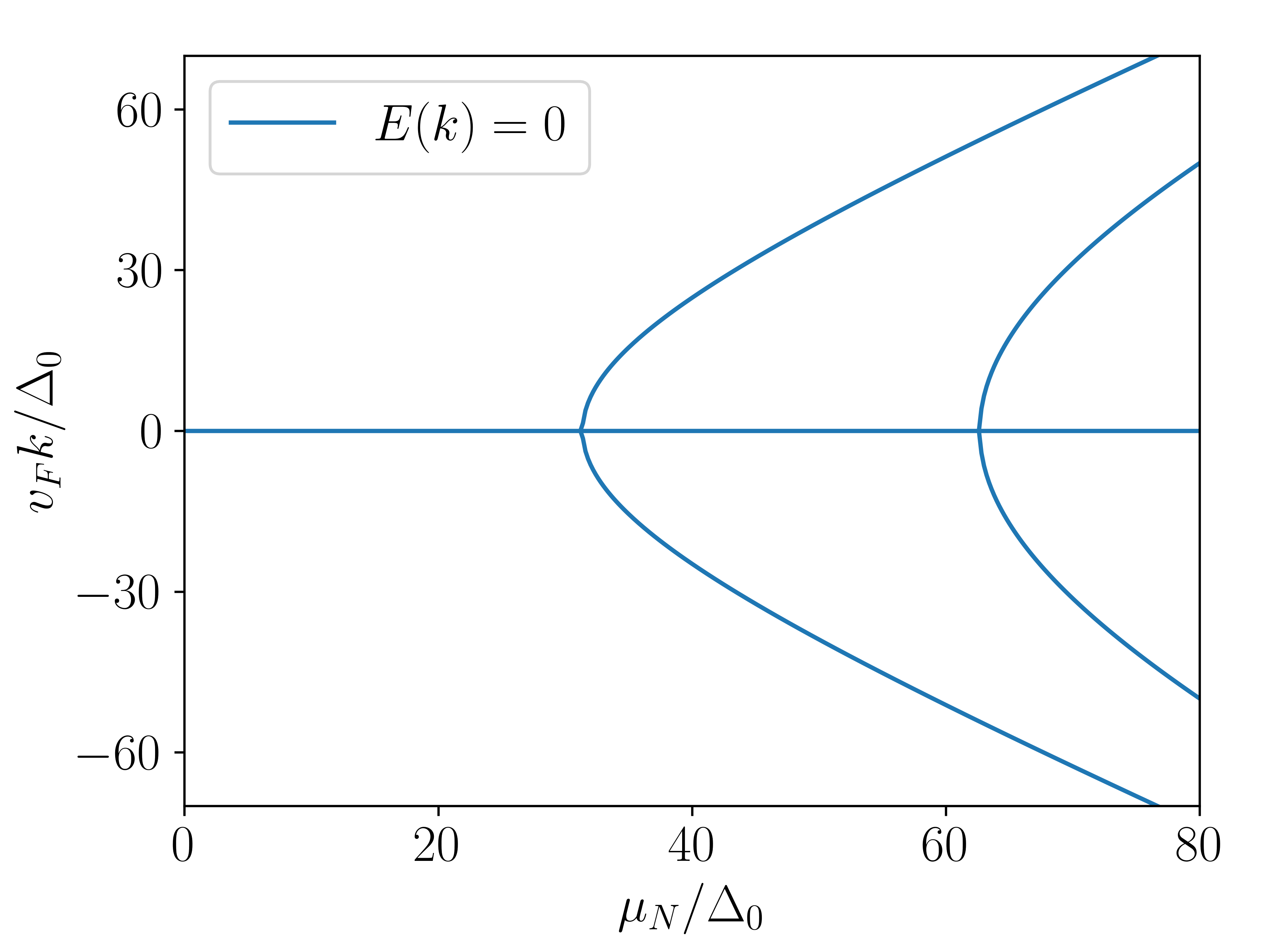}
    \caption{Centers of one-dimensional Dirac cones, determined by the values of $k$ for which $E(k)=0$, as a function of $\mu_N$ following from \eqref{eq:fukanehamiltonianwithk} for $W=0.1\,v_F/\Delta_0$, $\vec{M}=0$, $\mu_S = 100\Delta_0$. As $\mu_N$ grows, new Dirac cones emerge periodically, corresponding to instances where the effective velocity at $k=0$ vanishes.} 
    \label{fig:Dirac-cone-branches}
\end{figure}

We plot in Fig.~\ref{fig:Dirac-cones} the dispersion for the low-energy solutions of \eqref{eq:fukanehamiltonianwithk}, i.e. the energies $|E(k)|<\Delta_0$ at which an eigenmode can be found, for experimentally relevant values of the system parameters. In panel (a) for $\mu_N=0$ and a phase difference of $\varphi=\pi$, two linearly dispersing modes crossing at $E(k=0)=0$ are visible. These correspond to the familiar one-dimensional counter-propagating Majorana states. The slope is equal to the effective velocity $dE/dk|_{k=0}=v(\varphi=\pi) = v$. Away from $\varphi=\pi$, a gap opens, which allows Majorana zero modes to be bound to the Josephson vortices. 

As $\mu_N$ is increased in panel (b), the slope $v$ diminishes and we approach the point $\mu_{N,1}^{(0)}$ at which the effective velocity first vanishes. There, the non-linear nature of the dispersion near $k=0$ becomes apparent, corresponding to the higher-order corrections derived in Appendix \ref{app:higherordercorrections}. Increasing $\mu_N$ beyond $\mu_{N,1}^{(0)}$ in (c), two additional low-energy Dirac cones appear at finite momenta $k=\pm K$. These are gapless only at $\varphi=\pi$ as well. For growing $\mu_N$, more such Dirac cones periodically appear (see Fig.~\ref{fig:Dirac-cone-branches}), where the branching points $\mu_{N,j}^{(0)}$ all correspond to instances of vanishing effective velocity. In other words, after the $j$-th instance of the effective velocity at $k=0$ vanishing $\mu_{N,j}^{(0)}<\mu_N<\mu_{N,j+1}^{(0)}$, there are $2j+1$ Dirac cones at zero energy present. Note that the full spectrum of eigenstates was already obtained in a very similar analysis carried out for a graphene-based Josephson junction in Ref.~\onlinecite{titov_josephson_2006}, though the emergence of the extra Dirac cones was not noticed there. We also point out that the
presence of only odd numbers of Dirac cones resembles the behavior of the $\mathbb{Z}$ topological invariant introduced in Ref.~\onlinecite{Tewari_PRL2012} for a one-dimensional BDI system under the constraint of a fixed $\mathbb{Z}_2$ invariant.

Thus, in addition to the naive low-energy approximation missing the non-linear corrections to the dispersion, which become the leading order terms at $v=0$-points, we furthermore find an increased number of low-energy modes to periodically emerge with each zero-crossing of the effective velocity. Since these are also gapless at $\varphi=\pi$, one may expect them to lead to additional Andreev states bound to Josephson vortices in a magnetic field and thus to significantly alter the low-energy physics of the system. 
We conclude that in this parameter regime, the 1D Fu-Kane effective theory is not applicable. To confirm our suspicions regarding additional low-energy bound states, we introduce in the next section a different approach to finding the eigenmodes of \eqref{eq:hBdG-inmagneticfield} which does not rely on a limitation to the low-energy subspace.

In panels (d) and (e) of Fig.~\ref{fig:Dirac-cones}, it becomes apparent that the additional zero-energy Dirac degrees of freedom are not robust with respect to perturbations which break the $\tilde{U}_T$-symmetry. In real systems it is therefore to be expected that this mechanism results in additional near-zero energy Andreev states, rather than zero modes, even without taking into consideration hybridization between them (which will be discussed in the next section). In contrast, in panel (f) it can be seen that $M_x$ only acts like an additional phase difference, shifting the position at which the gap closes (see also Ref.~\onlinecite{reich_magnetization_2023}).

\section{Emergence of additional zero modes bound to Josephson vortices}
\subsection{Spectral matrix}

\begin{figure}
    \centering
    \includegraphics[width=\linewidth]{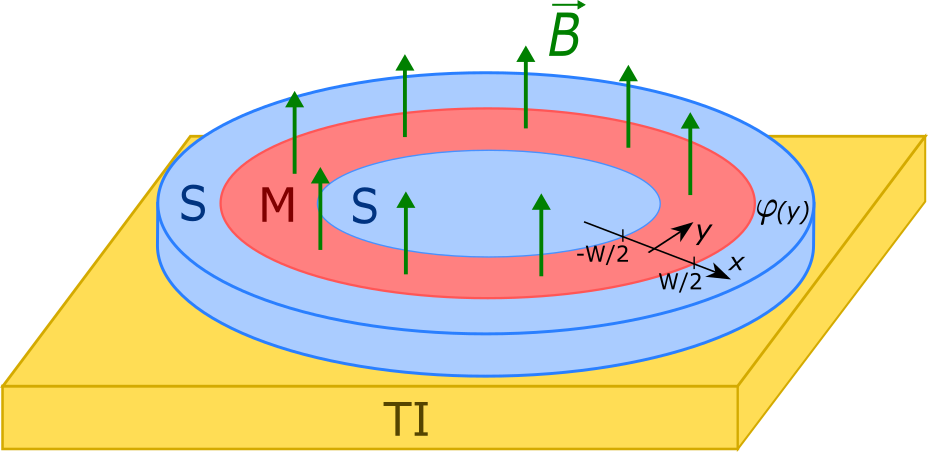}
    \caption{SMS junction on the surface of a 3D TI in an external magnetic field, as depicted in Fig.~\ref{fig:sketchSMSinmagneticfield}, closed into a Corbino ring geometry such that $y$ and $y+L$ correspond to the same point along the junction. The gauge is chosen for the phase of the inner superconductor to be 0 while the phase of the outer superconducting ring is given by $\varphi(y) = \varphi(y+L)(\text{mod}\,2\pi)$. The magnetic flux $\Phi = n\Phi_0$ is quantized such that $L = n l_B$, $n\in\mathbb{N}$.}
    \label{fig:sketch-corbino}
\end{figure}

Consider closing the Josephson junction into a Corbino ring geometry such that the system becomes periodic in $y$ (see Fig.~\ref{fig:sketch-corbino}). Consequently, the magnetic flux $\Phi$ through the junction can only take on values corresponding to integer multiples of the flux quantum, $\Phi = n\Phi_0$ with $n\in\mathbb{N}$, such that the length of the system $L$ is a multiple of the magnetic length $l_B$, $L = n l_B$, which ensures the single-valuedness of $\varphi$ (mod $2\pi$) at each point $y$. $n$ then also corresponds to the number of Josephson vortices (separated by one magnetic length from each other). For the sake of simplicity, we are going to assume a perfectly screened magnetic field and set the London penetration depth to zero, $\lambda_L = 0$. We expect the presented results to qualitatively hold true for small but non-zero $\lambda_L$ as well.

Taking advantage of the fact that the unitary transformation $U(\varphi) := e^{-i\varphi \tau_z/2}$ satisfies
\begin{align}
    U(\varphi)\tau_x U^{-1}(\varphi) = \cos\varphi\,\tau_x + \sin\varphi\,\tau_y,
\end{align}
in a complete basis of momentum eigenfunctions in $y$-direction $\braket{y|k_\ell} = \frac{1}{\sqrt{L}}e^{ik_\ell y/L}$,
the Hamiltonian \eqref{eq:hBdG-inmagneticfield} reads
\begin{align}
    &h_{\ell,\ell^\prime}:=\braket{k_\ell|h|k_{\ell^\prime}}\\
    &= \delta_{\ell,\ell^\prime}\Big\{\big[-iv_F\partial_x\sigma_x+\left(\frac{v_Fk_\ell}{L}+\frac{ev_F}{c}A_y(x)\tau_z\right)\sigma_y\notag\\
    &\qquad\quad- \mu(x)\big]\tau_z+ \vec{M}(x)\cdot\vec\sigma + \theta(-x-W/2)\Delta_0\tau_x\Big\} \notag\\ 
    &\qquad+\theta(x-W/2)\Delta_0\sum_{p}U_{\ell p}\tau_xU^\dagger_{p\ell^\prime}\notag
\end{align}
and the energy eigenvalue equation can be written as $\sum_{\ell^\prime}h_{\ell\ell^\prime}(x)\psi_{\ell^\prime}(x) = E\psi_\ell(x)$. To enforce antiperiodic boundary conditions corresponding to the Corbino geometry, it is chosen $k_\ell = (2\ell+1)\pi,\quad \ell \in \mathbb{Z}.$
The antiperiodicity results from a Berry phase $\pi$ that is picked up as one rotates to a local cartesian coordinate basis along the ring. For simplicity, we only consider $M_x=0$ in the following.

In each region, i.e. $x<-W/2\ (L)$ , $|x|<W/2\  (M)$, $x>W/2\  (R)$, we can explicitly find the linearly independent solutions to the respective eigenvalue equation. Since we are interested in Andreev bound states with $|E|<\Delta_0$, in the left and right region we choose only those two out of four that are zero at minus and plus infinity, respectively. The general solutions then are given by linear combinations of those
\begin{align}
    \psi_{L,\ell}(x;E) &= \Big(\psi^{(1)}_{L,\ell}(x;E), \psi^{(2)}_{L,\ell}(x;E)\Big)\begin{pmatrix}
        c^{(1)}_{L,\ell}(E)\\[.5em]c^{(2)}_{L,\ell}(E)
    \end{pmatrix}\notag\\
    &\equiv \Psi_{L,\ell}(x;E)C_{L,\ell}(E), \\
    \psi_{R,\ell}(x;E) &= \sum_{\ell^\prime}U_{\ell,\ell^\prime}\Big(\psi^{(1)}_{R,\ell^\prime}(x;E), \psi^{(2)}_{R,\ell^\prime}(x;E)\Big)\begin{pmatrix}
        c^{(1)}_{R,\ell^\prime}(E)\\[.5em]c^{(2)}_{R,\ell^\prime}(E)
    \end{pmatrix}\notag\\
    &\equiv \sum_{\ell^\prime}U_{\ell,\ell^\prime}\Psi_{R,\ell^\prime}(x;E)C_{R,\ell^\prime}(E)
\end{align}
where $\psi_{L/R,\ell}$ satisfy 
\begin{align}
    \Big(\big[-iv_F\partial_x\sigma_x+&\frac{v_Fk_\ell}{L}\sigma_y - \mu_S\big]\tau_z \label{eq:ev-sc}
    \\&\qquad+ \Delta_0\tau_x-E\Big)\psi^{(i)}_{L/R,\ell}(x;E) = 0 \notag
\end{align}
with $\psi^{(i)}_{L/R,\ell}(x\rightarrow \mp\infty;E) \rightarrow 0$. Even though $A_y(x>W/2) \neq A_y(x<-W/2)$, $\psi_{R,l}$ and $\psi_{L,l}$ are zero modes of the same operator, since for $x>W/2$ there is an additional contribution following from
\begin{align}
    \sum_p U_{\ell p}^\dagger U_{p\ell^\prime}k_p = \delta_{\ell,\ell^\prime}\left(k_\ell-n\pi\tau_z\right),
\end{align}
which cancels the $A_y$-term.
For the middle region it holds accordingly
\begin{align}
    \psi_{M,\ell}(x;E) &= \Big(\psi^{(1)}_{M,\ell}(x;E),\dots, \psi^{(4)}_{M,\ell}(x;E)\Big)\begin{pmatrix}
        c^{(1)}_{M,\ell}(E)\\\vdots\\c^{(4)}_{M,\ell}(E)
    \end{pmatrix}\notag\\ &\equiv \Psi_{M,\ell}(x;E)C_{M,\ell}(E)
\end{align}
with 
\begin{align}
    \Big(\big[-iv_F\partial_x\sigma_x+&\left(\frac{v_Fk_\ell}{L}+\frac{ev_F}{c}A_y(x)\tau_z\right)\sigma_y - \mu_N\big]\tau_z\notag\\
    &+ \vec{M}\cdot\vec\sigma-E\Big)\psi^{(i)}_{M,\ell}(x;E) = 0. \label{eq:ev-m}
\end{align}

In this form, we can write the matching conditions at the boundaries as
\begin{align}
    \Psi_{L,\ell}(-W/2;E)&C_{L,\ell}(E) = \Psi_{M,\ell}(-W/2;E)C_{M,\ell}(E), \notag\\[.3em] \sum_{\ell^\prime}U_{\ell,\ell^\prime}&\Psi_{R,\ell^\prime}(W/2;E)C_{R,\ell^\prime}(E)\\[-.7em] &\qquad= \Psi_{M,\ell}(W/2;E)C_{M,\ell}(E),\notag
\end{align}
from which follows
\begin{align}
    \sum_{\ell^\prime=-\infty}^\infty M_{\ell,\ell^\prime}(E)\begin{pmatrix}
        C_{R,\ell^\prime}(E) \\
        C_{L,\ell^\prime}(E)
    \end{pmatrix} = 0
\end{align}
with the 4$\times$4 matrices
\begin{align}
    M_{\ell,\ell^\prime}&=\Big(U_{\ell,\ell^\prime}\Psi_{R,\ell^\prime}({\scriptstyle \frac{W}{2}};E)\,,\\ &-\Psi_{M,\ell}({\scriptstyle \frac{W}{2}};E)\Psi^{-1}_{M,\ell}(-{\scriptstyle \frac{W}{2}};E)\Psi_{L,\ell}(-{\scriptstyle \frac{W}{2}};E)\,\delta_{\ell,\ell^\prime} \Big).\notag
\end{align}
Thus, the spectrum of the system is determined by the existence of zero modes of the infinite spectral matrix $M(E)$ with entries $M_{\ell,\ell^\prime}(E)$, which can be found as presented in Appendix \ref{app:entriesofspectralmatrix}.

Practically, we need to cut off $M(E)$ at some $|\ell|=\ell_c$ and argue that high momenta $k_{\ell>\ell_c}$ do not contribute to the low-energy physics. Due to this cut-off and finite numerical precision, we will not be able to find exact zero-modes of $M(E)$. The eigenvalue $\lambda_M(E)$ with the smallest magnitude will in general be non-zero. However, we observe that with an appropriately chosen and sufficiently large cut-off, values of the order of $|\lambda_M(E)|\lesssim 10^{-4}$, separated from the next largest eigenvalue by multiple orders of magnitude, can be achieved. We take it that, within the employed approximations, these can be interpreted as true zero-modes of $M(E)$.

In Fig.~\ref{fig:wavefunctions}, we are able to confirm that in the appropriate regime the results obtained via this method are in agreement with Ref.~\onlinecite{piasotski_topological_2024}.

\subsection{Spectral flow}

\begin{figure*}
    \centering
    \includegraphics[width=.75\linewidth]{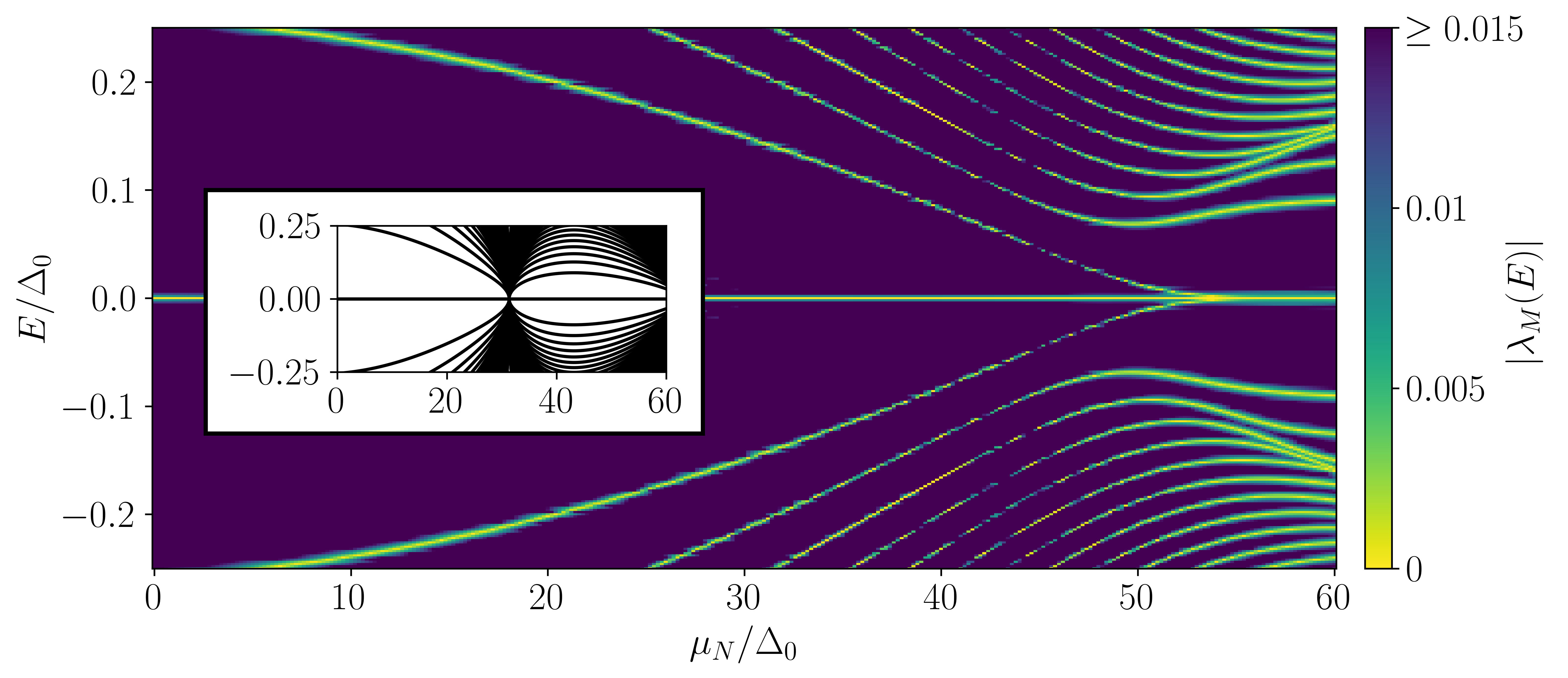}
    \caption{Bound state spectrum as a function of $\mu_N$ for $W=0.1\,v_F/\Delta_0$, $\vec{M}=0$, $\mu_S = 100\Delta_0$ and $L=l_B = 8\,v_F/\Delta_0$ ($\Phi = \Phi_0$). Shown in color is the absolute value of the lowest-lying eigenvalue $\lambda_M(E)$ of the spectral matrix $M(E)$ for the respective energy. Bright spots indicate the presence of an eigenmode. At $\mu_{N,1}^{(c)}\approx 55\Delta_0$, we observe that the first excited state merges with the zero-energy state, in accordance with the arguments presented in Sec.~\ref{sec:translationallyinvariant}, resulting in three-fold degenerate zero modes. The inset shows the prediction as derived in Ref.~\onlinecite{piasotski_topological_2024} according to the naive effective theory.}
    \label{fig:spectralflow}
    \vspace{.5cm}
    \includegraphics[width=.8\linewidth]{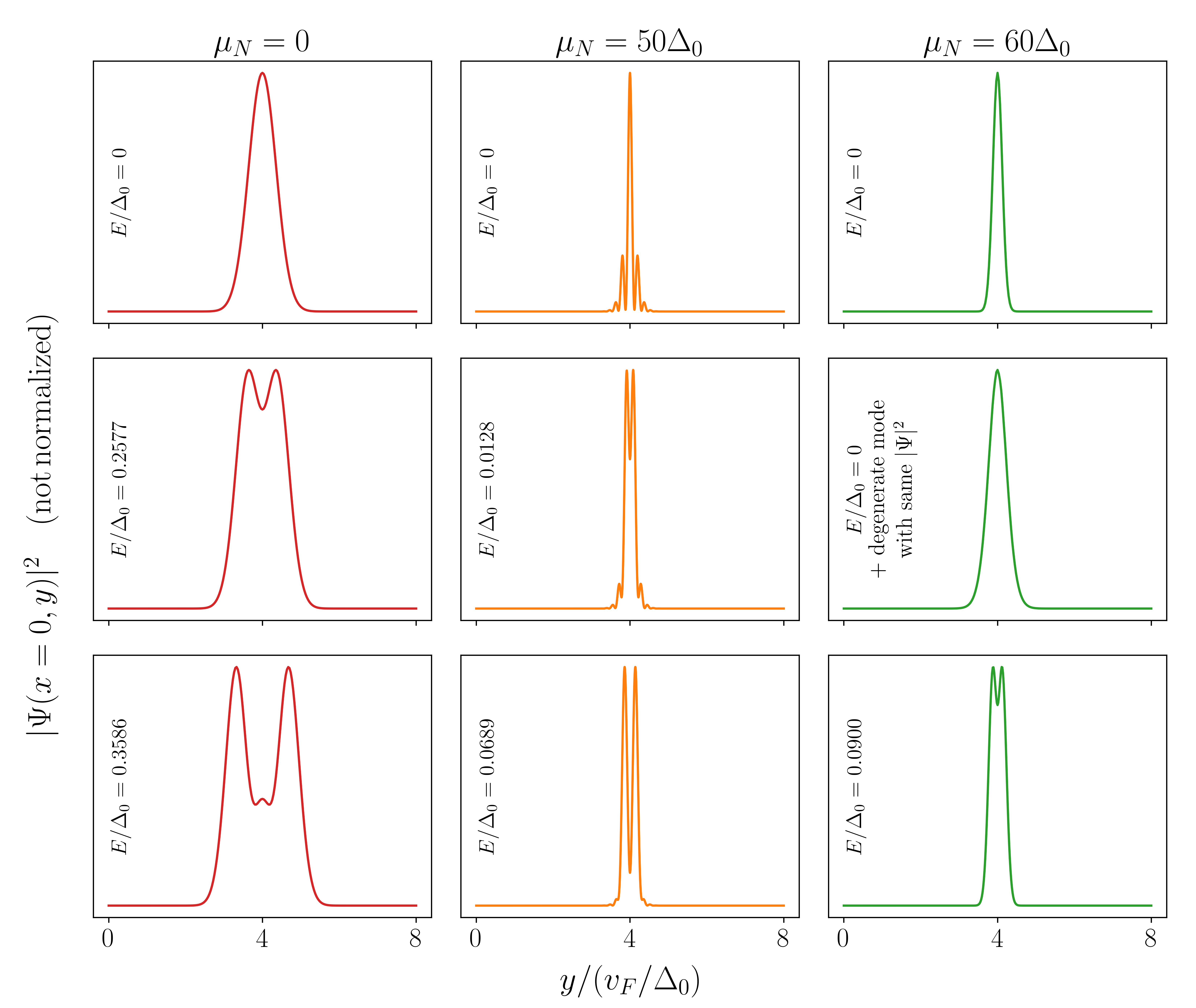}
    \caption{Probability densities of the lowest-energy bound states along slices at $x=0$ for the same parameters as in Fig.~\ref{fig:spectralflow} at $\mu_N=0$ (red), $\mu_N = 50\Delta_0$ (orange) and $\mu_N = 60\Delta_0$ (green), respectively. At $\mu_N=0$, the energies and shapes of the wave functions are in agreement with the predictions from Ref.~\onlinecite{piasotski_topological_2024} (their formulas give $E_1(\mu_N=0) \approx 0.2549\Delta_0$, $E_2(\mu_N=0) \approx 0.3605\Delta_0$). As $\mu_N$ grows close to $\mu_{N,1}^{(c)}$, the wave functions exhibit an oscillatory behavior due to the mixing with higher-$k$ modes. For $\mu_N = 60\Delta_0 >\mu_{N,1}^{(c)}$, the zero-energy subspace is threefold degenerate. The corresponding probability densities in Fourier space are shown in Fig.~\ref{fig:wavefunctions-kspace}. Furthermore, what previously was the second excited state takes on the role of the first excited state.}
    \label{fig:wavefunctions}
\end{figure*}

\begin{figure}
    \centering
    \includegraphics[width=\linewidth]{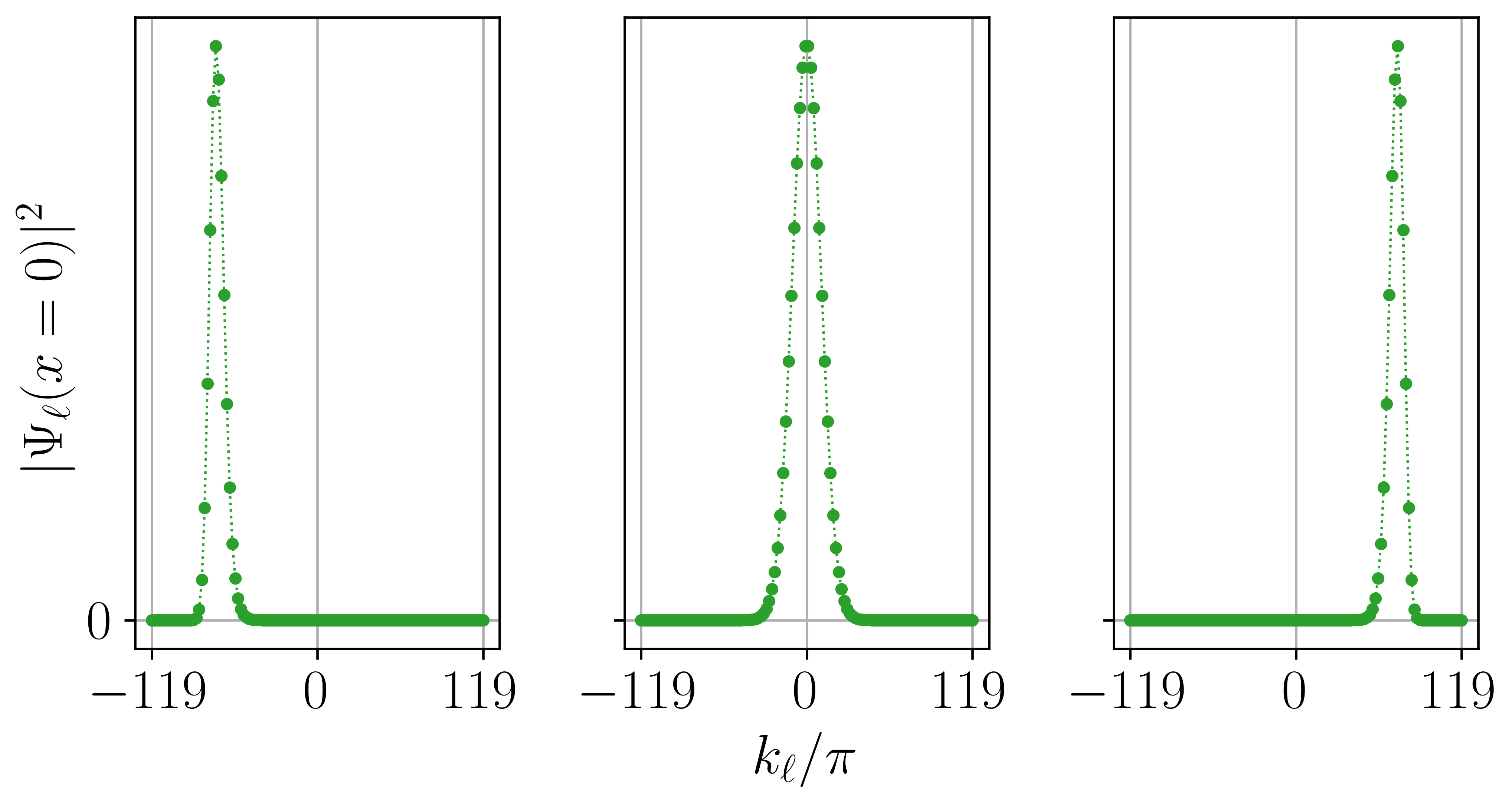}
    \caption{Probability densities of the three degenerate zero-energy states in Fourier space for a slice at $x=0$ at $\mu_N = 60\Delta_0$ for a cut-off $|k_{\ell_c}|/\pi = 119$. The corresponding pictures in real space are shown in Fig.~\ref{fig:wavefunctions}. The subspace only has support near three distinct values of $k_\ell$ symmetric around zero, as expected from Fig.~\ref{fig:Dirac-cone-branches}.}
    \label{fig:wavefunctions-kspace}
    \vspace{.5cm}
    \includegraphics[width=\linewidth]{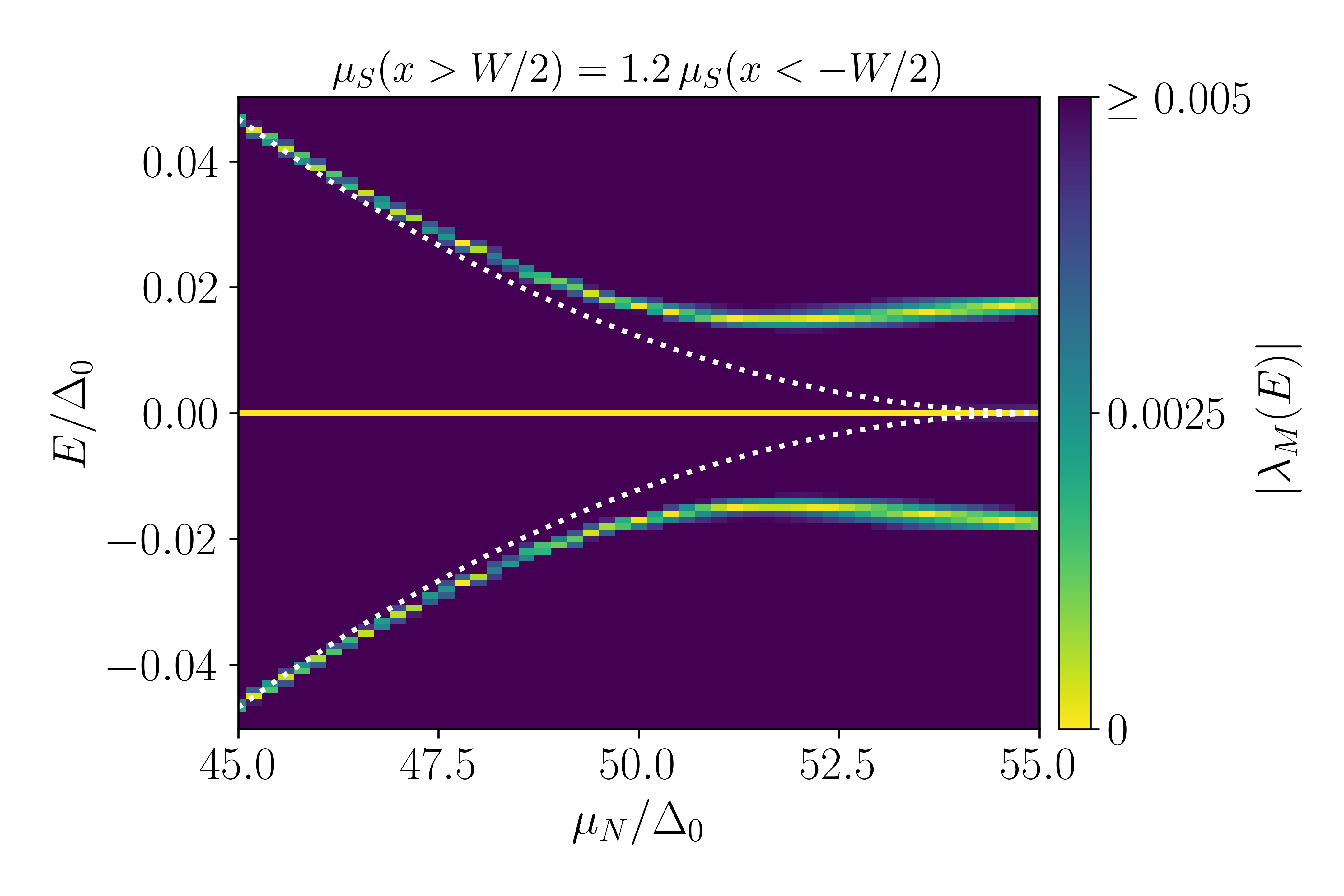}
    \caption{Spectral flow close to $\mu_{N,1}^{(c)}$ for the same parameters as in Fig.~\ref{fig:spectralflow} but with introduced asymmetry $\mu_S(x>W/2) = 1.2\mu_S(x<-W/2)$. A non-zero value of $M_y$, which also breaks the quasi-time-reversal symmetry $\tilde{U}_T$ defined in Sec.~\ref{sec:translationallyinvariant}, yields a similar picture. For comparison, the white dotted line indicates the previous result without asymmetry.}
    \label{fig:spectralflow_asymmetric}
\end{figure}

Employing the approach presented above, in the following we examine the energies of the CdGM states, tuning the system parameters such that the effective velocity of the Majorana modes goes through zero. We restrict ourselves to the case of $n=1$ flux quantum in the junction i.e. $n=1$ Josephson vortex near which all bound states are localized. The results can straightforwardly be generalized to $n>1$, with each vortex contributing the same bound state spectrum leading to a corresponding degeneracy, given the vortices are sufficiently far apart for no hybridization to take place due to overlap in position space.

In Fig.~\ref{fig:spectralflow} we plot the bound state spectrum as a function of $\mu_N$. In Ref.~\onlinecite{piasotski_topological_2024}, the authors found all bound states in the effective theory to condense around zero energy at the points at which the effective velocity vanishes (see inset of Fig.~\ref{fig:spectralflow}). While within the present approach an accumulation of bound states at low energies is still present, it occurs at a shifted value of $\mu_N$ compared to the effective theory. Furthermore, instead of all states, it is only the first excited state that reaches zero energy as $\mu_N$ is increased. The other ones remain at finite energy. The condensation of all bound states at zero energy within the effective theory can thus be attributed to its previously discussed loss of validity in the case of low effective velocities.

Notably, in accordance with the arguments in Sec.~\ref{sec:translationallyinvariant}, after merging with the zero-energy state at a value $\mu_N = \mu_{N,1}^{(c)}$, what previously was the first excited state remains at this energy and we obtain degenerate zero-energy states for $\mu_N>\mu_{N,1}^{(c)}$. Indeed, examining in Fig.~\ref{fig:wavefunctions-kspace} the shape of the three degenerate zero modes in $k$-space, they have support only near three distinct values of $k_\ell$ symmetric around zero. However, note that $55\Delta_0\approx\mu_{N,1}^{(c)}>\mu_{N,1}^{(0)}\approx 31\Delta_0$, i.e. the additional CdGM states do not manifest at the value at which we observed additional Dirac cones in Fig.~\ref{fig:Dirac-cone-branches}, but only at larger $\mu_N$ when there is already some separation between them in $k$-space. This can be understood as a consequence of hybridization, which is suppressed as the overlap in $k$-space decreases, in analogy to the usual hybridization of Majorana zero modes with overlapping wave functions in real space. This separation in momentum space is thus what allows multiple zero modes to be localized in the same vortex. 

It is important to note here that there are experimentally available topological insulators in which the underlying Dirac nature of the surface states, which is of course itself an effective low-energy theory, is valid even for energies $E\gtrsim 200\,$meV (see e.g. Ref.~\onlinecite{kushwaha2016sn}), such that the values for $\mu_{S/N}$ and $v_Fk$ employed here are well within realistic limits, assuming an induced superconducting gap $\Delta_0 \sim 0.5$\,meV. Note also, assuming the value for the Fermi velocity $v_F \simeq 4\,$eV\,\AA\ as determined in Ref.~\onlinecite{kushwaha2016sn}, that our choice for the width $W=0.1\,v_F/\Delta_0\sim 100\,$nm is compatible with recently conducted experiments.\cite{park2024corbino,park2026vortexparitycontrolleddiodeeffectcorbino} The substantial spatial fluctuations of the chemical potential in TI thin films reported in Ref.~\cite{brede2024characterizing} furthermore suggest that tuning $\mu_N$ to be sufficiently close to the Dirac point in the entire sample might not be straightforward.

In Fig.~\ref{fig:wavefunctions}, we additionally examine the shape of the corresponding probability densities in real space for the first few energy levels obtained from the spectral matrix-approach. For $\mu_N=0$, both the spectrum and the wave functions are in good agreement with Ref.~\onlinecite{piasotski_topological_2024}, where the Fu-Kane effective theory has been employed. As $\mu_N$ grows, so does the discrepancy between the two methods. Notably, close to the special point $\mu_{N,1}^{(c)}$, the probability densities obtain oscillatory components, which can be explained by the mixing with the new low-energy modes at higher values of $k$. After the zero-energy subspace has become threefold degenerate for $\mu_N>\mu_{N,1}^{(c)}$, one observes that the `new' first excited state, which previously was the second excited state, exhibits the same qualitative shape as the first excited state at $\mu_N=0$.

Breaking the mirror symmetry of the junction in Fig.~\ref{fig:spectralflow_asymmetric} finally reveals that the previously found newly emergent zero modes in this case instead remain at a small but finite energy. The additional bound states are thus in realistic systems no true zero modes, but ordinary CdGM states close to zero energy instead. 

In Refs.~\onlinecite{piasotski_topological_2024,laubscher2024detectionmajoranazeromodes}, the contribution of low-energy CdGM states to Josephson currents in such systems has been examined in response to some recently conducted experiments\cite{zhang2022ac,yue_signatures_2024,park2024corbino}, and microwave spectrocopy techniques have been proposed as a means to characterize the low-energy spectrum. Both these articles at least partly made use of the usually employed Fu-Kane low-energy theory. Our findings show that this approach does not necessarily yield reliable quantitative results for the interpretation of such experiments, that are sensitive to the fine details of the excitation spectrum, as it only produces the correct spectra for a limited range of parameters. Alternative approaches via 2D tight binding models\cite{abboud_signatures_2022,laubscher2024detectionmajoranazeromodes} might be able to resolve this problem, and it would be interesting to see whether they are indeed able to capture the additional zero crossings.

We would finally like to note that the qualitative behavior described here can be observed for a variety of chosen system parameters, including finite values of $M_z$. In the interest of brevity, we omit the corresponding plots and refer instead to Ref.~\onlinecite{Dissertation_Reich2025} for further examples.

\section{Conclusions and Outlook}

In this paper, we analyzed the low-energy excitation spectra of extended Josephson junctions fabricated on the surface of a 3D topological insulator. In particular, we examined a parameter regime in which the frequently employed Fu-Kane low-energy description, leading to a one-dimensional $2\times 2$ Dirac-like equation, is insufficient. 
We showed that the points, at which this effective theory loses validity, correspond to instances of vanishing effective (light) velocity and pointed out that such 
a scenario is experimentally realistic and achieved by, e.g., tuning the chemical potential of the normal/magnetic part of the junction.

Specifically, we revealed that the Fu-Kane effective theory misses a crucial aspect of the low-energy physics: in the translationally invariant situation, additional Dirac cones at zero energy, centered at non-zero momenta, emerge periodically anytime the effective velocity vanishes. If the junction is subjected to an external magnetic field, the number of zero modes bound to a Josephson vortex is therefore expected to periodically increase. In order to explicitly see this effect without the need for a low-energy `$k\cdot p$' approximation, we derived an alternative approach via an expansion in Fourier modes in a Corbino geometry of the junction. By means of this method, we were indeed able to see the growing discrepancy between the two approaches as the effective velocity diminishes as well as an increase of the number of zero modes bound to each Josephson vortex.

Although these additional zero modes are not robust against irregularities which break the inversion symmetry, they can be expected to significantly contribute to the low-energy physics and observables of real systems as low-lying CdGM states.

Possible future research directions lie in further investigations of the spectrum in the case of multiple Josephson vortices in instances, in which the overlap between neighboring CdGM states is significant, such that they form an effective one-dimensional vortex lattice. It would also be interesting to expand the methods employed here to take into account disorder and irregularities in the junction. Finally, establishing an explicit connection with the topological invariants introduced in Ref.~\onlinecite{Tewari_PRL2012} would be worthwhile.

\textit{Note added:} Recently, closely related results for edge modes in quantum anomalous Hall insulator-superconductor heterostructures have been independently
reported in the preprint Ref.~\onlinecite{yue2026chirality}, and dispersion relations displaying additional Dirac cones of the type we discuss in the present article have been presented for graphene-based Josephson junctions in the preprint Ref.~\onlinecite{varrica2026hybrid} based on the
results from Ref. \onlinecite{titov_josephson_2006}.

\section*{Acknowledgements}
We acknowledge support by the DFG grant SH 81 7-1.\\[.2em]

\section*{Data availability}
The data that support the findings of this article are openly available \cite{Reich2026_1000190009}.

\onecolumngrid
\begin{appendix}
\section{Derivation of the effective low-energy Hamiltonian and higher-order corrections \label{app:higherordercorrections}}

In order to first give a sketch of the general approach with which to obtain the (lowest-order) effective low-energy theory \eqref{eq:heff_anticomm} (for $M_x=M_y=0$), we separate the full 2D Hamiltonian \eqref{eq:hBdG-inmagneticfield} into two parts $h = h_0 + h_1$, taking $h_1 = -iv_F\tau_z\sigma_y\partial_y + \frac{ev_F}{c}A_y(x)\sigma_y$. The `transverse' part $h_0 = -iv_F\partial_x\sigma_x\tau_z- \mu(x)\tau_z+ M_z(x)\sigma_z + \Delta_0(x)\left(\cos\varphi(x,y)\tau_x + \sin\varphi(x,y)\tau_y\right)$ then only depends on $y$ parametrically through $\varphi(y)$ and the corresponding eigenvalue equation can be solved $h_0(x,\varphi(y))\xi_n(x,\varphi(y)) = \varepsilon_n(\varphi(y))\xi_n(x,\varphi(y))$ for each value of $\varphi$ with $\varepsilon_n(\varphi) = -\varepsilon_{-n}(\varphi)$. 

We next make the ansatz
\begin{align} \label{eq:fullexpansion_transversaleigenmodes}
    \psi(x,y) = \sum_n \alpha_n(y)\xi_n(x,\varphi(y))
\end{align}
for the solution to the full problem $(h_0+h_1)\psi = E\psi$. This can be understood as `gluing' together of one-dimensional slices at fixed $y$ in a way dictated by $h_1$. Since $\{\xi_n(x,\varphi(y))\}$ form a complete basis for every $y$, this is an exact procedure. The corresponding effective Hamiltonian acting on $\alpha_n(y)$ follows to be
\begin{align} \label{eq:heff-noanticomm}
    h_\text{eff}^{(m,n)}(y) = &-iv_F\braket{m|\tau_z\sigma_y|n}\partial_y- iv_F\braket{m|\tau_z\sigma_y|\partial_y n}+\frac{ev_F}{c}\braket{m|A_y(x)\sigma_y|n}+\varepsilon_n(\varphi(y))\delta_{m,n},
\end{align}
where $\braket{\cdot|\cdot|\cdot}$ only entails integration over $x$ and $\braket{x|n}=\xi_n(x,\varphi(y))$.

Rewriting
\begin{align}
    \braket{m|\tau_z\sigma_y|\partial_yn} = \frac{1}{2}\partial_y\braket{m|\tau_z\sigma_y|n} + \frac{1}{2}\Big(\braket{m|\tau_z\sigma_y|\partial_yn}-\braket{\partial_ym|\tau_z\sigma_y|n}\Big),
\end{align}
it follows
\begin{align} \label{eq:effhamiltonian-exact}
h_\text{eff}^{(m,n)}(y) = &-\frac{1}{2}\big\{v_{mn}(y),i\partial_y\big\} -\frac{1}{2}B_{mn}(y)+\braket{m|\Vec{R}_1(x)\cdot\Vec{T}|n}+\varepsilon_n(\vec{R}_0(y))\delta_{m,n},
\end{align}
where we defined
\begin{align}
    v_{mn}(y) := v_F\braket{m|\tau_z\sigma_y|n},\quad B_{mn}(y) := iv_F\Big(\braket{m|\tau_z\sigma_y|\partial_yn}-\braket{\partial_ym|\tau_z\sigma_y|n}\Big).
\end{align}

For $W\ll \xi$ the spectrum of $h_0$ for each value of $y$ consists only of two low-energy solutions $|\varepsilon_{\sigma=\pm}| \leq \Delta_0$. If then $E \ll \Delta_0$, it seems reasonable to assume that the full solution is well approximated by
\begin{align} \label{eq:twostateapprox}
    \psi(x,y) \approx \sum_{|n|\leq N}\alpha_n(y)\xi_n(x,\varphi(y)).
\end{align}
One thus obtains an effective $2\times 2$-Hamiltonian which is only dependent on $y$ given by
\begin{align}
   h_\text{eff}^{(\sigma,\sigma^\prime)}=-\frac{1}{2}\big\{v_{\sigma,\sigma^\prime}(y),i\partial_y\big\} -\frac{1}{2}B_{\sigma,\sigma^\prime}(y)+\frac{ev_F}{c}\braket{\sigma|A_y(x)\sigma_y|\sigma^\prime}+\varepsilon_n(\varphi(y))\delta_{\sigma,\sigma^\prime}.
\end{align}
The diagonal components of the effective velocity vanish
\begin{align}
    v_{\sigma,\sigma} = v_F\braket{\sigma|\tau_z\sigma_y|\sigma} = 0,
\end{align}
which follows from the quasi-time-reversal symmetry discussed in Sec.~\ref{sec:translationallyinvariant} of the main text. One may thus express the effective velocity as
\begin{align}
    v_{\sigma,\sigma^\prime}\equiv  v_\text{eff}^x(\varphi(y))\,\rho_x^{\sigma,\sigma^\prime} + v_\text{eff}^y(\varphi(y))\,\rho_y^{\sigma,\sigma^\prime},
\end{align}
where $\rho_i$ are Pauli matrices. By performing the gauge transformation 
\begin{align}
    \xi_\sigma \rightarrow \xi_\sigma e^{i\sigma \chi(y)/2},\quad \text{such that}\quad \braket{\xi_\sigma|\tau_z\sigma_y|\xi_{\sigma^\prime}}\in\mathbb{R},
\end{align}
we eliminate $v_\text{eff}^y$ and can thus define the effective velocity as $v_\text{eff}\equiv v_\text{eff}^x$. Note that this gauge transformation generates no additional contributions to $B_{\sigma,\sigma^\prime}$ since
\begin{align}
    \braket{\sigma|\tau_z\sigma_y|\partial_y{\sigma^\prime}}-\braket{\partial_y\sigma|\tau_z\sigma_y|{\sigma^\prime}}\quad  \rightarrow \quad &e^{-i(\sigma-\sigma^\prime)\chi/2}\left(\braket{\sigma|\tau_z\sigma_y|\partial_y{\sigma^\prime}}-\braket{\partial_y\sigma|\tau_z\sigma_y|{\sigma^\prime}}\right) \\
    &\hspace{1cm}+\frac{i}{2}\left[\partial_y\chi\right]\underbrace{\left(\sigma+\sigma^\prime\right)}_{=0\ \text{if}\ \sigma\neq\sigma^\prime}e^{-i(\sigma-\sigma^\prime)\chi/2}\underbrace{\braket{\sigma|\tau_z\sigma_y|{\sigma^\prime}}}_{=0\ \text{if}\ \sigma=\sigma^\prime}. \notag
\end{align}
From particle-hole symmetry follows furthermore
\begin{align}
    \braket{\sigma|\tau_z\sigma_y|\partial_y\sigma^\prime} = \braket{\partial_y(-\sigma^\prime)|\tau_z\sigma_y|(-\sigma)},
\end{align}
which allows us to write
\begin{align}
    B_{\sigma,\sigma^\prime}(y) = B_z(y)\rho_z^{\sigma,\sigma^\prime}.
\end{align}

Once again from particle-hole symmetry follows finally that $\braket{\sigma|\sigma_y|-\sigma} = 0$, such that, defining 
\begin{align}
\frac{ev_F}{c}\braket{\sigma|A_y(x)\sigma_y|{\sigma^\prime}} \equiv \rho_z^{\sigma,\sigma^\prime}\tilde{A}(y),
\end{align}
we arrive at an effective Hamiltonian of the form
\begin{align} \label{eq:heff-in2levelapprox}
h_\text{eff} = -\frac{1}{2}\big\{v_\text{eff}(y),\,i\partial_y\big\}\rho_x + \left(\varepsilon(y)+B_z(y) + \tilde{A}(y)\right)\rho_z
\end{align}
with $\varepsilon \equiv \varepsilon_+$.

In the main text we show that this approximation loses validity if at some point $v_\text{eff} = 0$. Let us therefore go back to the eigenvalue equation for the (exact) expression in Eq.~\eqref{eq:heff-noanticomm}, ignoring from now on the vector potential $A_y(x)$ for simplicity,
\begin{align} \label{eq:exacteigenvalueeqwithoutA}
    \sum_{n}\left[-iv_{mn}(y)\partial_y-\tilde{B}_{mn}(y)\right]\alpha_n(y) = (E-\varepsilon_m(y))\alpha_m(y),
\end{align}
where we defined $\tilde{B}_{mn}\equiv iv_F\braket{m|\tau_z\sigma_y|\partial_yn}$. The logic of our original two-level approximation works as follows: if the left-hand side, i.e. the contributions from $h_1$, are zero, the low-energy solutions read $E=\varepsilon_+$ with $\alpha_1 = 1$, $\alpha_{n\neq 1}=0$, and $E=\varepsilon_-=-\varepsilon_+$ with $\alpha_{-1}=1$, $\alpha_{n\neq -1}=0$. Then, we assume the corrections induced by $h_1$ to be small, and therefore the eigenstates to still have $\alpha_{n\neq \pm 1}\approx 0$, such that in the sum on the left-hand side we only have to take into account $n=\pm 1$. 

In other words, rewriting Eq.~\eqref{eq:exacteigenvalueeqwithoutA} as
\begin{align}\label{eq:iterativestartingpoint}
    &\Big[-iv_\text{eff}(y)\partial_y-\tilde{B}_{\sigma,-\sigma}(y)\Big]\alpha_{-\sigma}(y) - \Big[E-\varepsilon_\sigma(y)+\tilde{B}_{\sigma\sigma}(y)\Big]\alpha_\sigma(y) \notag\\
    &\qquad\qquad=-\sum_{n\neq \pm 1} \Big[-iv_{\sigma n}(y)\partial_y-\tilde{B}_{\sigma n}(y)\Big]\alpha_n(y)
\end{align}
with $\sigma = \pm 1$, we neglected the right-hand side of the equation since we assumed $\alpha_{n\neq \pm 1}\approx 0$ for $E\sim \varepsilon_\sigma$. 

In the following, we instead effectively integrate out the higher-energy contributions to take into account their higher-order effects on the low-energy subspace. To this end, one can rearrange Eq.~\eqref{eq:exacteigenvalueeqwithoutA}
\begin{align}
    \alpha_m(y) = \frac{1}{E-\varepsilon_m(y)+\tilde{B}_{mm}(y)}\sum_{n\neq m}\left[-iv_{mn}(y)\partial_y-\tilde{B}_{mn}(y)\right]\alpha_n(y) \label{eq:alpha_m(y)}
\end{align}
and iteratively plug this into the right-hand side of Eq.~\eqref{eq:iterativestartingpoint} (cf. Brillouin-Wigner perturbation theory). Carrying this procedure out up to second-order, one obtains an effective Hamiltonian of the form
\begin{align}
    h^\text{(corr)}_{\text{eff}} = &-\partial_y\left(\mathcal{A}(y)\partial_y\right)\rho_z+ \left(\varepsilon(y)+\mathcal{B}(y)\right)\rho_z \\&\qquad-\frac{1}{2}\left\{i\partial_y,v_\text{eff}(y)+\delta\tilde{v}_x(y)\right\}\rho_x-\frac{1}{2}\left\{i\partial_y,\delta\tilde{v}_y(y)\right\}\rho_y 
    -\frac{1}{2}\left\{i\partial_y,\delta\tilde{v}_0(y)\right\}\rho_0.\notag
\end{align} 
Most notably, disregarding the details of the individual parameters, in comparison to \eqref{eq:heff-in2levelapprox} a term $\sim \partial_y^2$ is generated, such that the Dirac nature of the Hamiltonian is lost, the dispersion becomes quadratic. Thus, if the leading contribution $v_\text{eff}$ becomes comparable to the presumably small corrections, the associated differential equation is governed by a higher-order derivative. The singular behavior we obtained above is thereby regularized. For further details not given here, see Ref.~\onlinecite{Dissertation_Reich2025}.

\section{Entries of the spectral matrix \label{app:entriesofspectralmatrix}}
In the superconducting regions, the eigenvalue equation \eqref{eq:ev-sc} can be rearranged to
\begin{align}
    v_F\partial_x \psi_{S} = \left[\frac{v_Fk_\ell}{L}\sigma_z+i\mu_S\sigma_x+\Delta_0\tau_y\sigma_x+iE\tau_z\sigma_x\right]\psi_{S}
\end{align}
which is solved by $\psi_{S}(x) = \xi_S e^{\kappa_S x/v_F}$ with $\kappa_S$ the eigenvalues of the matrix in the square brackets and $\xi_S$ the corresponding eigenvectors. For the left/right region we choose $\text{Re}(\kappa_S)\gtrless 0$, respectively.

Then
\begin{align}
    \Psi_{L,\ell}(-W/2;E) = \Big(\xi^{(1)}_{L,\ell}(E)e^{-\kappa^{(1)}_{L,\ell}(E)W/2},\quad \xi^{(2)}_{L,\ell}(E)e^{-\kappa^{(2)}_{L,\ell}(E)W/2}\Big)
\end{align}
and 
\begin{align}
    \Psi_{R,\ell}(W/2;E) = \Big(\xi^{(1)}_{R,\ell}(E)e^{\kappa^{(1)}_{R,\ell}(E)W/2},\quad \xi^{(2)}_{R,\ell}(E)e^{\kappa^{(2)}_{R,\ell}(E)W/2}\Big).
\end{align}

Due to the $x$-dependence of $A_y$ in the middle region, the solutions are more involved. Since the Hamiltonian here is diagonal in particle-hole space, we can consider $\tau_z=\pm 1$ separately. The eigenvalue equation \eqref{eq:ev-m} (for $M_x=0$) can then be written as
\begin{align}
    \left(\pm v_F(-i\partial_x)\sigma_x+\frac{v_F\pi}{l_BW}\left(x+\frac{W}{2}+\frac{l_B W}{v_F\pi}M_y\pm \frac{l_BW}{\pi}\frac{k_\ell}{L}\right)\sigma_y+M_z\sigma_z\right)\psi^{(\pm)}_{M,\ell}(x) = (E\pm\mu_N)\psi^{(\pm)}_{M,\ell}(x).
\end{align}
Defining the dimensionless coordinate
\begin{align}
    \Tilde{x}_\pm := \sqrt{\frac{\pi}{l_B W}}\left(x+\frac{W}{2}+\frac{l_B W}{v_F\pi}M_y\pm \frac{l_BW}{\pi}\frac{k_\ell}{L}\right) \quad \Rightarrow\quad \partial_x = \sqrt{\frac{\pi}{l_B W }}\partial_{\Tilde{x}_\pm}
\end{align}
and introducing in each case the harmonic oscillator ladder operators
\begin{align}
    \Tilde{x}_\pm = \frac{1}{\sqrt{2}}(a^\dagger+a),\quad -i\partial_{\Tilde{x}_\pm} = \frac{i}{\sqrt{2}}(a^\dagger-a)
\end{align}
we find
\begin{align}
    \left[\pm i(a^\dagger\sigma_\mp - a \sigma_\pm)+\Tilde{M}_z\sigma_z\right]\psi^{(\pm)}_{M,\ell}(\Tilde{x}) = \Tilde{E}_\pm \psi^{(\pm)}_{M,\ell}(\Tilde{x})
\end{align}
with $\Tilde{M}_z = \sqrt{\frac{l_BW}{2\pi v_F^2}}M_z$, $\Tilde{E}_\pm = \sqrt{\frac{l_BW}{2\pi v_F^2}}(E\pm \mu_N)$ and $\sigma_\pm = \frac{1}{2}(\sigma_x\pm i\sigma_y)$.

The four linearly independent solutions read (up to normalization)
\begin{gather}
    \psi^{(1)}_{M,\ell}(x;E) = \left(i \phi^{K_+-1}_{0,\ell}(\tilde x_+)\ket{\uparrow_\sigma}+(\Tilde{M}_z-\Tilde{E}_+)\phi^{K_++1}_{1,\ell}(\tilde x_+)\ket{\downarrow_\sigma}\right)\ket{\uparrow_\tau}, \\
    \psi^{(2)}_{M,\ell}(x;E) = \left((\Tilde{M}_z+\Tilde{E}_+) \phi^{K_+-1}_{1,\ell}(\tilde x_+)\ket{\uparrow_\sigma}-i\phi^{K_++1}_{0,\ell}(\tilde x_+)\ket{\downarrow_\sigma}\right)\ket{\uparrow_\tau},\\
    \psi^{(3)}_{M,\ell}(x;E) = \left(-i \phi^{K_-+1}_{0,\ell}(\tilde x_-)\ket{\uparrow_\sigma}+(\Tilde{M}_z-\Tilde{E}_-)\phi^{K_--1}_{1,\ell}(\tilde x_-)\ket{\downarrow_\sigma}\right)\ket{\downarrow_\tau},\\
    \psi^{(4)}_{M,\ell}(x;E) = \left((\Tilde{M}_z+\Tilde{E}_-) \phi^{K_-+1}_{1,\ell}(\tilde x_-)\ket{\uparrow_\sigma}+i\phi^{K_--1}_{0,\ell}(\tilde x_-)\ket{\downarrow_\sigma}\right)\ket{\downarrow_\tau}
\end{gather}
with $\ket{{\mathbin\uparrow\hspace{-.5em}\downarrow}_{\{\sigma,\tau\}}}$ denoting the eigenvectors of $\sigma_z$ and $\tau_z$, respectively. We defined here the generalized harmonic oscillator eigenfunctions (see e.g. Ref.~\onlinecite{sen_physical_2024})
\begin{gather}
    \phi_{0,\ell}^K(x) = {}_{1}F_1\left(\frac{1-K}{4},\frac{1}{2},x^2\right)e^{-x^2/2},\\
     \phi_{1,\ell}^K(x) = {}_1F_1\left(\frac{3-K}{4},\frac{3}{2},x^2\right)\,x\,e^{-x^2/2}
\end{gather}
with $K_\pm = \Tilde{E}_\pm^2-\Tilde{M}_z^2$ and $_1F_1(a,b,z)$ the confluent hypergeometric function of the first kind.

Finally, it is
\begin{align}
    U_{\ell,\ell^\prime} &= \frac{1}{L}\int_0^Ldy\,e^{-i\varphi(y)\tau_z/2}e^{-i(k_\ell-k_{\ell^\prime})y/L} = e^{-i\varphi_0\tau_z/2}\int_0^1d\Tilde{y}\,e^{-i\pi (n\tau_z+2(\ell-\ell^\prime))\Tilde{y}}\\
    &= e^{-i\varphi_0\tau_z/2}\sigma_0\begin{cases}
        \begin{pmatrix}
            0 & 0 \\ 0 & 1
        \end{pmatrix},\quad &n=2(\ell-\ell^\prime),\\[1em]
        \begin{pmatrix}
            1 & 0 \\ 0 & 0
        \end{pmatrix},\quad &n=-2(\ell-\ell^\prime),\\
        \frac{i}{\pi}((-1)^n-1)\begin{pmatrix}
            (2(\ell-\ell^\prime)+n)^{-1} & 0 \\ 0 & (2(\ell-\ell^\prime)-n)^{-1}
        \end{pmatrix},\quad &\text{else}.
    \end{cases} \notag
\end{align}
where we remember $L=n\cdot l_B$.

Note that for the case of even $n$, each Fourier mode is pairwise coupled to only two others, since the non-diagonal contributions are given by $U_{l,l^\prime} \sim \delta_{n,2|l-l^\prime|}$ (in contrast to odd $n$, where the coupling between Fourier modes extends infinitely far $\sim [2(\ell-\ell^\prime)\pm n]^{-1}$). When employing a hard cut-off, the modes with $\ell = \pm \ell_c$ thus only couple to one mode each. This allows for unphysical non-trivial solutions (regardless of the parameters) where the contributions from the subspace $\{-\ell_c,-\ell_c+2,\dots,\ell_c\}$ vanish, although these kinds of eigenstates cannot be present in the infinite matrix-limit. To combat this artifact of the cut-off procedure in the case of even $n$, one can e.g. artificially introduce an additional matrix element between $+\ell_c$ and $-\ell_c$, such that there exists no decoupled subspace. With the cut-off being chosen large enough for these modes not to significantly contribute to the low-energy physics, one then obtains physically sensible results which are independent of this fictitious coupling.

\end{appendix}

\twocolumngrid
\bibliography{references}

\end{document}